\documentclass[twocolumn,openany,openright,amsmath,amssymb,superscriptaddress]{revtex4-2}

\usepackage{float}
\usepackage{latexsym} 
\usepackage{amsmath,amsthm}
\usepackage{ifpdf}
\usepackage{subfig}
\usepackage{soul}
\usepackage{dcolumn}
\usepackage{bm}
\usepackage{wrapfig}
\usepackage[dvipsnames]{xcolor}
\usepackage{color}
\usepackage{hyperref}
\usepackage{soul}

\usepackage{graphicx}

\begin{document}
\newcommand{\umlaut}[1]{\ddot{\textrm{#1}}}
\newcommand{\beq}{\begin{equation}}
\newcommand{\eeq}{\end{equation}}
\newcommand{\barr}{\begin{eqnarray}}
\newcommand{\earr}{\end{eqnarray}}
\newcommand{\bseq}{\begin{subequations}}
\newcommand{\eseq}{\end{subequations}}
\newcommand{\oper}[1]{\hat{#1}}
\newcommand{\adj}[1]{\oper{#1}^{\dagger}}
\newcommand{\ann}{\hat{a}}
\newcommand{\creat}{\hat{a}^{\dagger}}
\newcommand{\commutator}[2]{[\oper{#1},\oper{#2}]}
\newcommand{\bra}[1]{\langle #1|}
\newcommand{\ket}[1]{|#1\rangle}
\newcommand{\outerP}[2]{|#1\rangle\langle#2|}
\newcommand{\expectation}[3]{\langle #1|#2|#3\rangle}
\newcommand{\proiettore}[1]{\ket{#1}\bra{#1}}
\newcommand{\closure}[1]{\sum_{#1}\proiettore{#1}=1}
\newcommand{\closureInt}[1]{\int d#1\proiettore{#1}=1}
\newcommand{\braket}[2]{\langle #1|#2\rangle}
\newcommand{\vett}[1]{\textbf{#1}}
\newcommand{\uvett}[1]{\hat{\textbf{#1}}}
\newcommand{\derivata}[2]{\frac{\partial #1}{\partial #2}}

\newcommand{\creak}[1]{\hat{a}^{\dagger}_{#1}}
\newcommand{\annak}[1]{\hat{a}_{#1}}
\newcommand{\crebk}[1]{\hat{b}^{\dagger}_{#1}}
\newcommand{\annbk}[1]{\hat{b}_{#1}}

\newcommand{\crek}[2]{\hat{#1}^{\dagger}_{#2}}
\newcommand{\annk}[2]{\hat{#1}_{#2}}
\newcommand{\red}[1]{{\color{red}{#1}}}
\newcommand{\blue}[1]{{\color{blue}{#1}}}
\newcommand{\upp}{\uppercase}

\title{2D Weyl Materials in the Presence of Constant Magnetic Fields}
\author{Yaraslau Tamashevich}
\affiliation{Faculty of Engineering and Natural Sciences, Tampere University, Tampere, Finland}
\author{Leone Di Mauro Villari}
\affiliation{Department of Physics and Astronomy, University of Manchester, Manchester M13 9PL, UK}
\author{Marco Ornigotti}
\affiliation{Faculty of Engineering and Natural Sciences, Tampere University, Tampere, Finland}

\begin{abstract}
In this work we investigate the effect of a constant external, or artificial, magnetic field on the nonlinear response of 2D Weyl materials.  We calculate the Landau Levels for tilted cones in 2D Weyl materials by treating the tilting in a perturbative manner, and employ perturbation theory to calculate the tilting-induced correction to the magnetic field induced Landau spectrum. We then calculate the induced current as a function of the tilting coefficients and extract the correspondent nonlinear signal. Then, we analyze how changing tilting parameter affects nonlinear signal. Our findings show the possibility of achieving a significant tunability of the nonlinear response, by suitably engineering the orientation and degree of tilt of Dirac cones in 2D Weyl materials.
\end{abstract}

\maketitle

\section{\label{sec:Intoduction}Introduction}
In the last eighteen years, after the discovery of graphene \cite{ng,ng1}, the study of two dimensional pseudo-relativistic fermions has dominated the condensed matter arena. The study of graphene in an uniform magnetic field also contributed to this boom with the discovery of the anomalous, \emph{half-integer} quantum Hall effect \cite{Zhang}. The basic theory stems from the quantisation of cyclotron orbits in a uniform magnetic field, the Landau quantisation (LQ). The carriers can only occupy orbits with discrete, equidistant energy values, called the Landau levels (LL) \cite{LL}. LQ plays an important role in the electronic properties of materials. It is directly responsible for diamagnestism and, at strong magnetic field, leads to oscillation of the magnetic susceptiblity and conductivity  known as De Haas-Van Alphen \cite{ha} and Shubnikov-de Haas  \cite{sa1,sa2} effects respectively. 

Over the past decade, several theoretical and experimental works delat with the problem of characterising the behaviour of Weyl semimetals in the presence of magnetic fields, including describing the effect of tilting on the LL spectrum both semiclassically \cite{extra1}, and with a full relativistic quantum treatment \cite{extra2}. These results have been applied to characterise the the effect of the tilt on LL spectroscopy, and revealed salient features on the evolution of the absorption spectrum of 2D Dirac fermions upon tilt, such as the insurgence of non-dipolar transitions, as a consequence of the tilt \cite{extra3,extra4}. The effect of tilting on LL has also been studied in 3D Weyl materials, where similar results have been found \cite{extra5}.

Despite the great interest this topic has attracted, especially on 2D materials, no study on the simultaneous effect of cone tilting and LL on the nonlinear optical response of such materials has been carried out, to the best of our knowledge.  For this reason, in this paper we focus our attention on LL of tilted, type-II, Weyl materials (WM). Type-II WM were first postulated in 2015 \cite{sg}. Contrary to Type-I WM which have straight cones at the nodal point, and preserve Lorentz invariance, they are tilted and with broken Lorentz invariance.

Type II cones typically occur when Type I Weyl nodes are tilted enough, along some specific direction, that a Lifschitz transition occurs and the system acquires a finite density of states at the Weyl node. Type II fermions, either Dirac or Weyl, have been found in a variety of materials such as semimetals, TMDs (PtTe$_2$, WTe$_2$) \cite{my}, LaAlO$_3$/ LaNiO$_3$/ LaAlO$_3$ quantum wells \cite{tt} and PdTe$_2$ superconductors \cite{nj}. The electronic properties of these materials have been extensively studied in the last years \cite{zzc,kke,bdb,gts}. The nonlinear optics has also recently started to attract some attention \cite{mgl,xh,tvo}. 

In this work we focus on two dimensional materials with type II Weyl fermions (WF),which have been theoretically explored with increasing attention in the last years \cite{lgb,zx,hnn,gz,cs,Lin}. Although their conclusive experimental observation is still lacking they are not unrealistic. In fact, a promising platform for the experimental realisation of such materials is, for example, the organic compound $\alpha$-(BEDT-TTF)$_2$I3, a quasi-2D conductor which supports WF \cite{bhs,kt,tk}. In particular, we study the nonlinear optical response of the Landau levels in type-II WM. Although a full relativistic and nonperturbative approach, following Ref. \cite{extra2}, could be employed to calculate the effect of the tilting on LL, we instead opt for a perturbative approach, which allows us to write a set of coupled mode equations for the time-dependent population coefficients and calculate the induced current as a function of such coefficients and the correspondent nonlinear spectrum, directly in the lab frame, rather than in the boosted frame. This allows for a more direct, and experiment-friendly approach to the problem. Notice, however, that treating the tilting perturbatively does not compromise the nature of the Dirac cones, and, ultimately, the nature of the considered material. Type II materials, in fact, are characterised by tilted cones, regardless if their tilting is strong or weak. A more complete framework could be setup by using the approach described in detail in Ref. \cite{extra2}.

 Our results clearly show, how it is possible to control and enhance the nonlinear response of 2D Weyl materials using an external, or artificial (generated, for example, through bending or strain \cite{graphGauge}), magnetic field. We, in fact, show, how a suitable combination between magnetic field control and material engineering, in the form of control of the degree of tilt of Dirac cones, can lead to efficient generation of high harmonics, up to order 50, and beyond. This could lead to significant applications in optics and photonics. The possibility to control the nonlinear optical response of such materials by controlling both the applied magnetic field and the degree of tilt of Dirac cones in such materials, in fact, could lead to reconfigurable, broadband, efficient frequency converters. On the other hand, the same device could also be used for sensing applications, essentially exploiting the fact, that external perturbation can change either the local crystalline structure of Weyl materials (thus amounting to an overall change of the intensity of the applied magnetic field, for example), or they can induce anisotropies, that could change the degree of tilting, thus allowing the use of the intrinsic anisotropic nature of the nonlinear optical response of 2D Weyl materials \cite{ourPRB}. In both cases, analysing the effects of such perturbations on the nonlinear optical response of 2D Weyl materials might prove insightful for sensing applications.

This paper is organised as follows: In section II we introduce the Hamiltonian for tilted two dimensional Weyl materials in the presence of an external gauge field. In section III we compute the tilt induced perturbation on Landau levels. In section IV and V we focus on the nonlinear optical response by considering the interaction of the Landau level with the impinging electric field. Finally, conclusions are drawn in Sect. VI.
\section{Tilted Hamiltonian for 2D Weyl Materials}\label{sec:hamiltonian}
2D Weyl material in the presence of an external electromagnetic field (described by the vector potential $\vett{A}(t)$) and a gauge field (described by the U(1) gauge potential $\vett{A}^{(g)}$) is given by
\beq \label{eq1}
%
\hat{H}=\sum_{i=1}^2v_ip_i\hat{\sigma}_i+\Delta\hat{\sigma}_{3}+\vett{a}\cdot\vett{p}\,\hat{\sigma}_0,
\eeq
where $\hat{\sigma}_k$ are Pauli matrices (with $\hat{\sigma}_0=\mathbb{I}$ is the two-dimensional identity matrix), $\vett{a}=(a_x,a_y)$ is the tilting vector, $v_{x,y}$ is the (anisotropic) Fermi velocity along the $x$- and $y$-direction, respectively, $\Delta$ is a staggered potential, which accounts for the gap between the valence and conduction bands, and $\vett{p}$ is the minimally coupled kinetic momentum, which takes into account both the impinging electromagnetic field and the external (or artificial) magnetic field, i.e.,
\beq\label{eq2}
p_{\mu}=k_{\mu}+\frac{e}{c}A_{\mu}^{(g)}+\frac{e}{c}A_{\mu},
\eeq
where $k_{x,y}$ are the Cartesian components of the momentum. For later convenience, we also define the minimally coupled \emph{magnetic} momentum $\pi_{\mu}=k_{\mu}+\frac{e}{c}A_{\mu}^{(g)}$, which will be useful in constructing the unperturbed and tilted LL for 2D Weyl materials. 
\begin{figure}[!t]
    \centering
    \includegraphics[width=\linewidth]{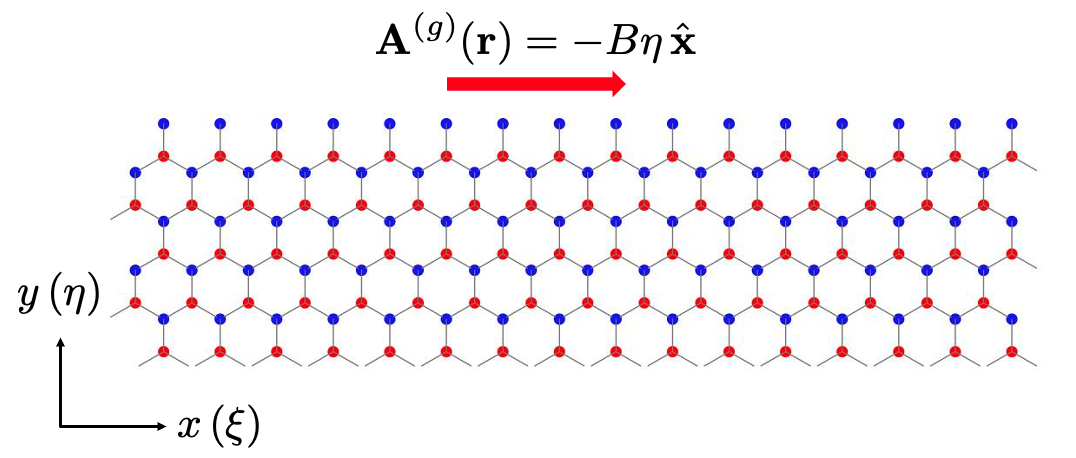}
    \caption{Schematic representation of the ribbon geometry for a 2D Weyl material. The ribbon is assumed to be infinitely extended along the $x-$direction, so that traslational symmetry is unbroken along it. The gauge potential $\vett{A}^{(g)}$, giving rise to the out-of-plane magnetic field, is oriented along the $x$-direction, so that its magnitude can be proportional to $y$, thus allowing to decouple Landau quantisation from the calculation of the effective band structure. The figure shows a regular honeycomb lattice, where the two different colours (blue and red) of the lattice sites represent the different atomic species characterising a Weyl materials. Both unscaled ($\{x,y\}$) and scaled ($\{\xi,\eta\}$) reference frames are shown for convenience.}
    \label{geometry}
\end{figure}
In general, the presence of the vector potential $\vett{A}^{(g)}$ in Eq. \eqref{eq2} breaks traslational symmetry, rendering Bloch theorem, and the correspondent tight-binding approximation, invalid. To overcome this problem, we assume to consider a ribbon geometry, as the one depicted in Fig. \ref{geometry}, where $x-$ is assumed to be infinitely extended, while $y-$ has a finite extension. In this way, $k_x$ is still a good quantum number, as traslational invariance is not broken along the $x-$direction, and we can calculate an effective $y$-dependent band structure. If, moreover, we choose the gauge potential $\vett{A}^{(g)}$ in such a way, that  $\vett{A}^{(g)}\propto y$, we can apply standard Landau quantisation without breaking traslational symmetry in the $x-$direction. 
\begin{figure}[!t]
    \centering
    \includegraphics[width=\linewidth]{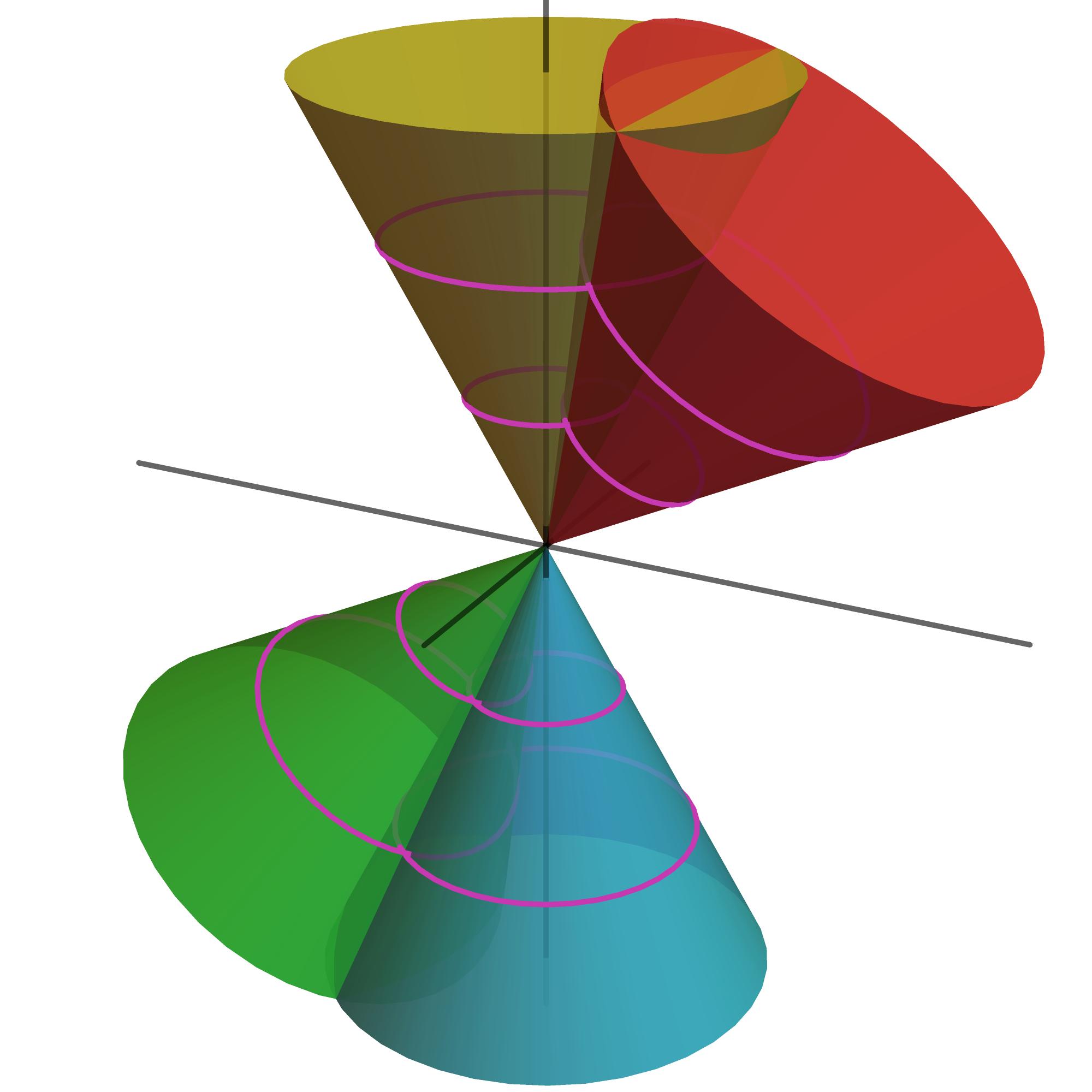}
    \caption{Pictorial representation of untilted vs. tilted cones in 2D Weyl materials. The blue and orange cones represent the valence and conduction band (respectively) of an untilted system ($\vett{a}=0$ in Eq. \eqref{eq1}), while the green and red cones depict tilt-induced perturbation of the band structure in the vicinity of Dirac point. The purple lines on each cone represent the LL appearing in that band due to the magnetic field.}
    \label{fig:two_cones}
\end{figure}

The Hamiltonian above can be split into two contributions, one of which only contains the magnetic field through the magnetic momentum $\pi_{\mu}$, and the other containing only information about the tilt, though the parameters $a_{x,y}$. Before doing that, however, since in general
$v_x\neq v_y$, it is first convenient to rescale both the momentum and the gauge fields $\vett{A}$ and $\vett{A}^{(g)}$ by introducing the effective Fermi velocity $v=\sqrt{v_xv_y}$ and, consequently, the scaled momentum coordinates $k_{\xi}=\sqrt{v_x/v_y}\,k_x $, $k_{\eta}=\sqrt{v_y/v_x}\,k_y$, and the correspondent scaled gauge field components $A_{\xi}=\sqrt{v_x/v_y}\,A_x$ and $A_{\eta}=\sqrt{v_y/v_x}A_y$, so that the following transformation between magnetic momenta holds
\beq
v_x\pi_x\pm iv_y\pi_y\hspace{0.5cm}\rightarrow\hspace{0.5cm}v(\pi_{\xi}\pm i\pi_{\eta}),
\eeq
meaning that then, the components of the magnetic momentum in the scaled frame are given by
$\pi_{\xi} = \sqrt{v_x/v_y}\,\pi_x$ and $\pi_{\eta} = \sqrt{v_y/v_x}\,\pi_y$.

\hfill\\
After doing this, we can then rewrite the Hamiltonian in Eq. \eqref{eq1} as  $\hat{H}\equiv \hat{H}_0+\hat{H}_{tilt}$, where $\hat{H}_0$ only depends on the gauge fields (and, in particular, on the applied magnetic field) but not on the tilting parameters, and its explicit expression is given by
\beq\label{hamH0}
\hat{H}_0=v\left[\sum_{i=1}^2\pi_i\hat{\sigma}_i+\frac{\Delta}{v}\hat{\sigma}_3\right]+\frac{ev}{c}\sum_{i=1}^2A_i\hat{\sigma}_i.
\eeq
The tilt Hamiltonian, on the other hand, contains dependence on the tilting parameter $\tau_k=a_k/v_k$ and reads
\beq\label{hamTilt}
\hat{H}_{tilt}=\sum_{i=1}^2\left[\tau_i\left(\pi_i+\frac{e}{c}A_i\right)\right]\hat{\sigma}_0,
\eeq
where now the convention $1\rightarrow\xi$ and $2\rightarrow\eta$ has been implicitly assumed, and $\tau_{\xi,\eta}\equiv\tau_{x,y}$.

A pictorial representation of the band structure of 2D Weyl materials in the vicinity of one of their Dirac points is given in Fig. \ref{fig:two_cones}. In particular, the band structure given by the Hamiltonian $\hat{H}_0$ in Eq. \eqref{hamH0} imposes Landau levels on a straight Dirac cone (orange and blue cones in Fig. \ref{fig:two_cones}), while the inclusion of the tilting Hamiltonian of Eq. \eqref{hamTilt} tilts the whole structure, i.e., Dirac cone plus Landau levels (green and red cones in Fig. \ref{fig:two_cones}).
%
\section{Landau Quantisation of Tilted Cones}\label{sec:tilted_landau_levels}
In this section we shall see how to perform the LQ for the Hamiltonian (\ref{eq1}). To do that, we will adopt a perturbative approach, where we first quantise $\hat{H}_0$, which will give rise to the usual LL, and then include the tilting perturbatively, by promoting $\tau_x$ and $\tau_y$ to perturbative parameters. As a result of this operation, we will show how the tilting de-facto introduces new transition matrix elements between the valence and conduction band LL. 
\subsection{Standard Landau Quantisation}
To start with, let us first consider $\hat{H}_0$ and solve the eigenvalue problem
\beq\label{eq6}
\hat{H}_0\ket{\phi_n}=E_n\ket{\phi_n},
\eeq
in the presence of a constant magnetic field. Since we are considering a ribbon geometry (see Fig. \ref{geometry}), we can exploit traslational symmetry along the $x-$direction to write the eigenstate $\ket{\phi_n}$ as
\beq
\ket{\phi_n}=\int\,dk_{\xi}\,e^{ik_{\xi}\xi}\,\ket{\psi_n(\eta;k_{\xi})},
\eeq
where $\ket{\psi_n(k_{\xi};\eta)}\equiv\ket{\psi_n}$, emphasises that the eigenstate $\ket{\psi_n}$ is function of the momentum $k_{\xi}$ and depends parametrically on $\eta$. Notice, moreover, that even in the absence of a magnetic field, i.e., for $\vett{A}^{(g)}=0$, the above Ansatz is still valid, and allows for the calculation of an effective band structure, which will depend parametrically on $\eta$.

As described above, the magnetic field is inserted through minimal coupling, by the substitution $
\vett{k}\rightarrow\boldsymbol\pi=\vett{k}+\frac{e}{c}\vett{A}^{(g)}$,
where $\vett{A}^{(g)}$ is the vector potential corresponding to the applied magnetic field. To make calculations easier, we can assume to work in the Landau gauge, where the vector potential can be chosen as 
\beq\label{eq7}
\vett{A}^{(g)}=-B\eta\uvett{x}.
\eeq
Please notice, that at this level of analysis, we don't make any assumption on the nature of the magnetic field. It could be an actual, external, magnetic field, or it could emerge as a consequence of bending or straining the material lattice, as detailed in Ref. \citenum{graphGauge}. Our formalism and calculations are insensitive to the physical origin of the magnetic field, and for this reason we will not specify one.

The easiest way to solve the above eigenvalue problem \eqref{eq6}, is to transform $\hat{H}_0$ in terms of creation and annihilation operators of the harmonic oscillator. To do that, let us first observe that $[\pi_{\xi},\pi_{\eta}] = -i\left(\frac{eB}{c}\right)$ \cite{landau}.
This suggests to take $q_{\xi}=(c/eB)\pi_{\eta}$ as a viable canonically conjugated generalised coordinate to the generalised momentum $\pi_{\xi}$. This means, that $\pi_{\xi}$ and $\pi_{\eta}$ are canonically conjugated variables, and we can therefore associate creation and annihilation operators to them, as per standard Quantum Mechanics \cite{messiah}, i.e.,
\bseq
\begin{align}
\pi_{\xi}&=\frac{1}{\sqrt{2}L_B}\left(\hat{a}^{\dagger}-\hat{a}\right),\\
\pi_{\eta}&=\frac{i}{\sqrt{2}L_B}\left(\hat{a}^{\dagger}+\hat{a}\right),
\end{align}
\eseq
where $[\hat{a},\hat{a}^{\dagger}]=1$, and $L_B=\sqrt{\hbar/eB}$ is the characteristic magnetic length. Substituting this into the scaled $\hat{H}_0$, and introducing the quantities $\lambda\equiv\Delta L_B/\sqrt{2}v$ and $\omega_c=\sqrt{2} v/L_B$, we have
\beq\label{eq13}
\hat{H}_0=\omega_c\left(\begin{array}{cc}
\lambda& \hat{a}^{\dagger}\\
\hat{a} & -\lambda
\end{array}\right)
\eeq

The eigenvalues and eigenstates of the Hamiltonian. in Eq. \eqref{eq13} can be then readily calculated by substituting its expression in to Eq. \eqref{eq6},so that the eigenvalue spectrum is finally given by
\beq\label{eq10}
\varepsilon_n=s_n\sqrt{\lambda^2+|n|},
\eeq
where $|n|$ accounts for the Landau level, and $s_n=\text{Sign}(n)\equiv\pm$ accounts for valence ($-$) and conduction ($+$) bands, respectively. The (normalised) Landau spinor corresponding to the eigenstates of $\hat{H}_0$ then reads
\beq
\ket{\psi_n}=\frac{1}{\sqrt{1+\gamma_n^2}}\left(\begin{array}{c}
\phi_{|n|}\\
\gamma_n\phi_{|n|-1}
\end{array}\right),
\eeq
where $\gamma_n = (\varepsilon_n-\lambda)/\sqrt{|n|}$, and $\phi_n(x)$ is a harmonic oscillator eigenstate (Hermite-Gaussian functions in 1D) \cite{landau}, and the prescription, that $\phi_{-1}=0$ is enforced. Notice, moreover, that for $n=0$ we have
\bseq
\begin{align}
\varepsilon_0&=\lambda,\\
\ket{\psi_0}&=\left(\begin{array}{c}
\phi_0\\
0
\end{array}\right),
\end{align}
\eseq
and that we need to exclude the eigenvalue $\varepsilon_0^-=-\lambda$, because for this case $\gamma_0^-\rightarrow\infty$, and therefore $\ket{\psi_0^-}\rightarrow 0$.
\subsection{Introducing the tilting}\label{sect3b}
We now consider the effect of the tilt on both Landau levels and eigenstates. We first rewrite $\hat{H}_{tilt}$ in terms of creation and annihilation operators, following the prescription above, to obtain
\beq\label{eq13}
\hat{H}_{tilt}=\omega_c\left[\tau_x(\hat{a}^{\dagger}-\hat{a})+i\tau_y(\hat{a}^{\dagger}+\hat{a})\right]\mathbb{I},
\eeq
where $\tau_i\equiv a_i/2v_i$ is the the tilting in the $i$-direction, and constitutes our perturbation parameter. To calculate the effect of the tilting on the Landau levels and eigenstates of $\hat{H}$ in Eq. \eqref{eq1}, we employ first-order perturbation theory, using $\tau$ as the perturbation parameter. Without loss of generality, we can assume that the tilt only happens along the $x$-direction, so that $\tau_y=0$ and $\tau\equiv\tau_x$, and the calculations are easier. This assumption essentially means, that we consider the situation in which the vector potential (generating the external out-of-plane magnetic field) is aligned to the tilting direction. A more general solution, with vector potential and tilting misaligned is readily obtained by employing a perturbation theory with two perturbative parameters $\tau_{x,y}$, or by simply applying a rotation operator to both the vector potential and the eigenstates of $\hat{H}_0$, but it is out the scope of this paper.

First, we then expand the eigenstates and eigenvalues of the tilted system in terms of the perturbative parameter $\tau$, up to order $\mathcal{O}(\tau^2)$, i.e., $\ket{\psi_n}=\ket{\psi_n^{(0)}}+\tau\ket{\psi_n^{(1)}}+\mathcal{O}(\tau^2)$, and $\varepsilon_n=\varepsilon_n^{(0)}+\tau\varepsilon_n^{(1)}+\mathcal{O}(\tau^2)$, and solve the complete eigenvalue problem up to the same order, thus obtaining
\bseq\label{eq14}
\begin{align}
\left(\hat{H}_0 + \tau\hat{V}\right)\left(\ket{\psi_n^{(0)}}+\tau\ket{\psi_n^{(1)}}\right)\\ \nonumber =\left(\varepsilon_n^{(0)}+\tau\varepsilon_n^{(1)}\right)&\left(\ket{\psi_n^{(0)}}+\tau\ket{\psi_n^{(1)}}\right),
\end{align}
\eseq
where $\hat{V}\equiv\hat{H}_{tilt}/(\tau\omega_c)$. The zero-order solution has been given above (and corresponds to the untilted cones). At first order in $\tau$ we have instead
\beq\label{eq23}
\hat{H}_0\ket{\psi_n^{(1)}}+\hat{V}\ket{\psi_n^{(0)}}=\varepsilon_n^{(0)}\ket{\psi_n^{(1)}}+\varepsilon_n^{(1)}\ket{\psi_n^{(0)}},
\eeq
Notice, moreover, that the normalisation condition of $\ket{\psi_n}$ imposes that
$\braket{\psi_n^{(0)}}{\psi_n^{(1)}}=0$. With this information, we can then calculate the correction to the eigenvalues by projecting Eq. \eqref{eq23} onto $\bra{\psi_n^{(0)}}$, obtaining
\beq
\varepsilon_n^{(1)}=\expectation{\psi_n^{(0)}}{\hat{V}}{\psi_n^{(0)}}.
\eeq
It is not hard to show,that
$\varepsilon_n^{(1)}=0.$
Substituting this result into Eq. \eqref{eq23} and applying standard methods of perturbation theory we get
\beq\label{eq27}
\ket{\psi_n^{(1)}}=\sum_{m\neq n}\frac{\expectation{\psi_m^{(0)}}{\hat{V}}{\psi_n^{(0)}}}{\varepsilon_n^{(0)}-\varepsilon_m^{(0)}}\ket{\psi_k^{(0)}}.
\eeq
Using the (normalised) expression of $\hat{V}$ through $\hat{H}_{tilt}$ and the expressions of the untilted eigenstates we get
\beq\label{eq18}
\expectation{\psi_m^{(0)}}{\hat{V}}{\psi_n^{(0)}}=\alpha_{n,n+1}\delta_{m,n+1}-\alpha_{n,n-1}\delta_{m,n-1},
\eeq
where
\bseq\label{eq19}
\begin{align}
\alpha_{n,n+1}&=\frac{\sqrt{|n|+1}+\gamma_n\gamma_{n+1}\sqrt{|n|}}{\sqrt{(1+\gamma_n^2)(1+\gamma_{n+1}^2)}},\\
\alpha_{n,n-1}&=\frac{\sqrt{|n|}+\gamma_n\gamma_{n-1}\sqrt{|n|-1}}{\sqrt{(1+\gamma_n^2)(1+\gamma_{n-1}^2)}}.
\end{align}
\eseq
Substituting this result into Eq. \eqref{eq27}, we get the following form for the perturbed eigenstates
\beq\label{eq20}
\ket{\psi_n^{(1)}}=N\left(\begin{array}{c}
A_{n,n+1}\,\phi_{|n|+1}+A_{n,n-1}\,\phi_{|n|-1}\\
\gamma_{n+1}A_{n,n+1}\,\phi_{|n|}+\gamma_{n-1}A_{n,n-1}\,\phi_{|n|-2}
\end{array}\right),
\eeq
with $N=1/\sqrt{1+\gamma_n^2}$ and 
\beq
A_{n,n\pm 1}=\frac{\sqrt{1+\gamma_n^2}}{\varepsilon_n^{(0)}-\varepsilon_{n\pm 1}^{(0)}}\alpha_{n,n\pm 1}.
\eeq
This is the first result of our work. At first order in perturbation theory, the tilting only affects the eigenstates $\ket{\psi_n}$ but not the energies of the LL. Since, as it can be seen in the equation above, the perturbative correction to the Landau eigenstates contains terms proportional to $\phi_{|n|\pm 1}$ and $\phi_{|n|-2}$, the expected impact of this perturbation in the interaction of tilted cones with an impinging electormagnetic pulse would be to modify the selection rules for the allowed transitions, thus inserting more excitation and decay channels, then the untilted case.
\section{Coupled Mode Equations for the Interaction Hamiltonian}\label{sec:coupled_mode_equation}
\begin{figure}
    \centering
    \includegraphics[width=0.8\linewidth]{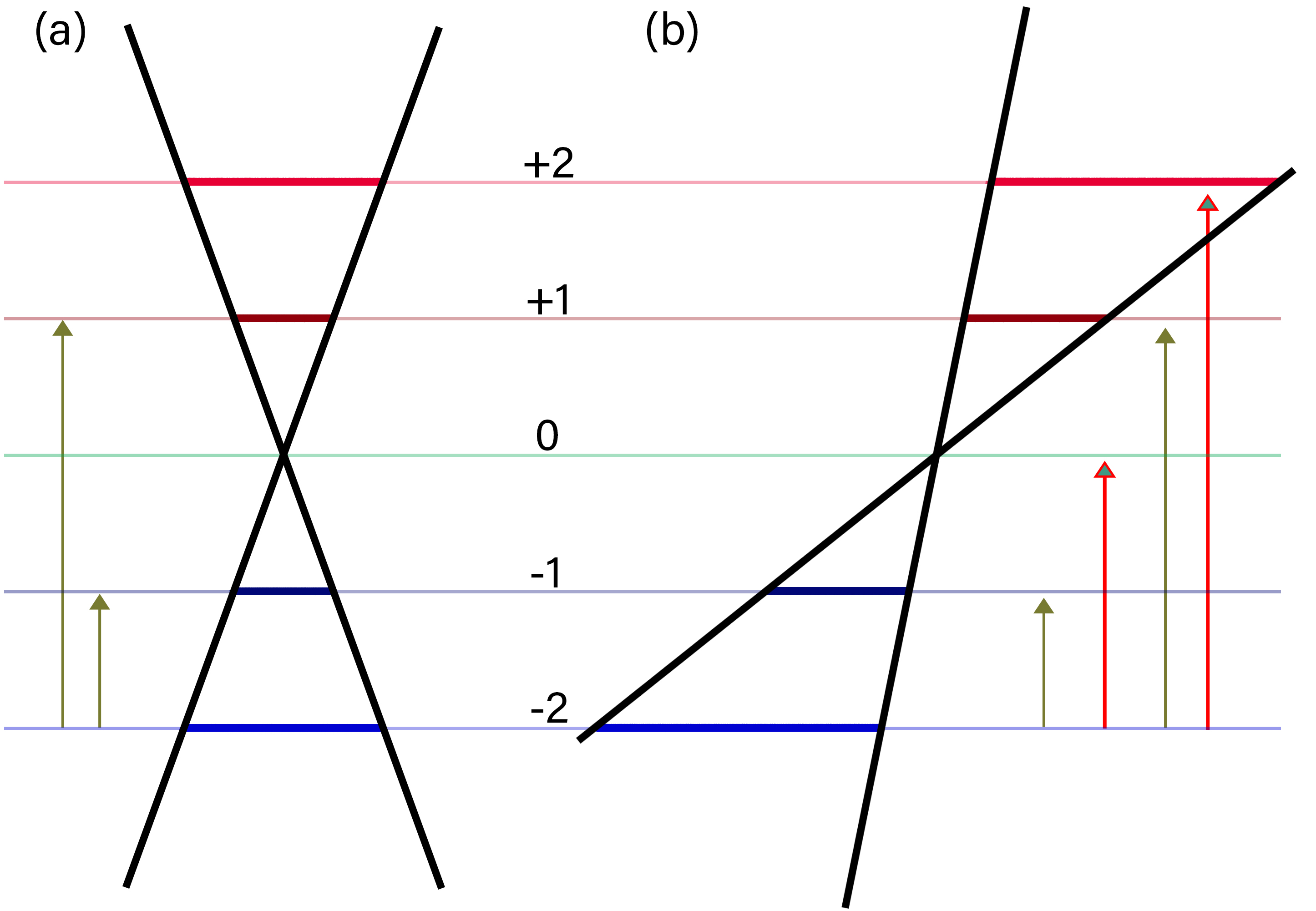}
    \caption{Band structure of LL in the vicinity of the Dirac points, for the case of (a) untilted and (b) tilted cones. As it can be see, in panel (a), the selection rules dictate, for example, that only the transitions represented by gray lines, i.e., $\ket{\psi_{-2}}\rightarrow\ket{\psi_{-1}}$ and $\ket{\psi_{-2}}\rightarrow\ket{\psi_{1}}$, are possible (assuming, that $\ket{\psi_{-2}}$ is the only initially populated level). When the cone is tilted [panel (b)], however, the red transitions, i.e., $\ket{\psi_{-2}}\rightarrow\ket{\psi_0}$ and $\ket{\psi_{-2}}\rightarrow\ket{\psi_2}$ become allowed. This is possible due to the tilting extending the selection rules to encompass also $\Delta|n|=2$ as a viable option.Please notice, that in going from panel (a) to panel (b), while the valence and conduction bands are being tilted, the energy levels are not, as they are, at the first order in perturbation theory, unaffected by the tilting. This, ultimately, is one of the reasons for the appearance of extra selection rules.}
    \label{fig:landau_levels}
\end{figure}
In this section, we consider the action of an impinging, pulsed electromagnetic field and study its interaction with a system presenting tilted cones, i.e., a Weyl material. We continue working within the minimal coupling framework, and simply add a second gauge field to the picture, this time representing the external pulse impinging on the Weyl material, by employing the substitution
$\boldsymbol\pi\rightarrow\textbf{p}=\boldsymbol\pi+(e/c)\vett{A}(t)$. The interaction Hamiltonian, comprising, as it can be seen from Eqs. \eqref{hamH0} and \eqref{hamTilt}, both tilt-independent and tilt-dependent terms, then reads
\beq
\hat{H}_{light}=\frac{ev}{c}\left[\sum_{i=1}^2A_i\hat{\sigma}_i+2\tau\left(A_{\xi}+i A_{\eta}\right)\mathbb{I}\right].
\eeq
Without loss of generality, we can assume the impinging optical pulse to be polarised along the $\xi$-direction (i.e., along the $x$-direction), so that we can set $A_{\eta}=0$ and write
\beq\label{eq23}
A_{\xi}\equiv A(t)= E_0\tau \,e^{-\frac{(t-t_0)^2}{\Delta t^2}}\cos(\omega_L t),
\eeq
where $E_0$ is the amplitude of the impinging electric field, $\Delta t$ the pulse duration, and $\omega_L$ is the pulse central frequency. Notice, that since in the previous section we have assumed the Dirac cone to be tilted along the $x$-direction (i.e., we set $\tau_y=0$), the assumption made above essentially corresponds to the case in which the impinging field is polarised along the tilting direction.

With the light-matter Hamiltonian at hand, we can then cast the time-dependent Dirac equation to the following form
\beq
i\frac{\partial}{\partial t}\ket{\Psi(t)}=\left(\hat{H}_0+\hat{H}_{tilt}+\hat{H}_{light}(t)\right)\ket{\Psi(t)}.
\eeq
To solve it, we expand the real solution in the instantaneous eigenstates calculated above, i.e.,
\beq\label{eq37}
\ket{\Psi(t)}=\sum_n\left[c_n^+(t)e^{-i\omega_n t}\ket{\psi_n^+}+c_n^-(t)e^{i\omega_n t}\ket{\psi_n^-}\right],
\eeq
where $\ket{\psi_n}=\ket{\psi_n^{(0)}}+\tau\ket{\psi_n^{(1)}}$, $\omega_n=|\varepsilon_n|=\sqrt{\lambda^2+|n|}$. and $\sum_n\left(|c_n^+|^2+|c_n^-|^2\right)=1$. Substituting this Ansatz in the equation above and remembering that $\ket{\psi_n}$ are (perturbative) eigenstates of $\hat{H}_0+\hat{H}_{tilt}$, we get the following set of coupled mode equations for the time-dependent coefficients
\bseq
\label{eq:coupled_mode_equations}
\begin{align}
i\,\dot{c}_m^+ &= \sum_n\Big[\expectation{\psi_m^+}{\hat{H}_{light}(t)}{\psi_n^+}e^{-i(\omega_n-\omega_m)t}\,c_n^+\nonumber\\
&+\expectation{\psi_m^+}{\hat{H}_{light}(t)}{\psi_n^-}e^{i(\omega_n+\omega_m)t}\,c_n^-\Big],\\
i\,\dot{c}_m^- &= \sum_n\Big[\expectation{\psi_m^-}{\hat{H}_{light}(t)}{\psi_n^+}e^{-i(\omega_n+\omega_m)t}\,c_n^+\nonumber\\
&+\expectation{\psi_m^-}{\hat{H}_{light}(t)}{\psi_n^-}e^{i(\omega_n-\omega_m)t}\,c_n^-\Big].
\end{align}
\eseq

To solve these coupled mode equations, one first needs to write down the matrix elements $\expectation{\psi_m^{\pm}}{\hat{H}_{light}(t)}{\psi_n^{\pm}}$. To do so, let us assume, that the carrier frequency of the impinging pulse is resonant with the transition $\ket{\psi_{-1}}\rightarrow\ket{\psi_0}$ (corresponding to the transition between the zero energy state and the first LL in valence band), and then set $\omega_L=|\omega_0-\omega_1|=\sqrt{\lambda^2+1}-\lambda$. 

In the untilted case, selection rules only allow transitions that obey $\Delta|n|=1$ \cite{Falko}, which corresponds to choosing matrix elements of the form $\expectation{\psi_{m,\pm}^{(0)}}{\hat{H}_{light}}{\psi_{n,\pm}^{(0)}}\simeq \delta_{|m|,|n|+1}+\delta_{|m|,|n|-1}$.

In the presence of tilt, however, the selection rule above is not valid anymore, and new selection rules appear, due to the fact, that the matrix elements of the interaction Hamiltonian calculated with respect to the perturbed eigenstates read
\beq
\expectation{\psi_{m,\pm}^{(1)}}{\hat{H}_{light}}{\psi_{n,\pm}^{(1)}}\simeq \delta_{|m|,|n|+2}+\delta_{|m|,|n|-2}.
\eeq
Combining this extra set of selection rules with those of the usual, untilted, case, we can write the tilt-induced selection rules as follows
\beq
0<\Delta|n|\leq 2.
\eeq
This is the second result of our work. The tilt in Dirac cones introduces an extra set of selection rules for light-matter interaction, which essentially originates from the eigenstate mixing imposed by the tilt itself [see Eq. \eqref{eq20}].
\begin{figure}
    \centering
    \includegraphics[width=0.8\linewidth]{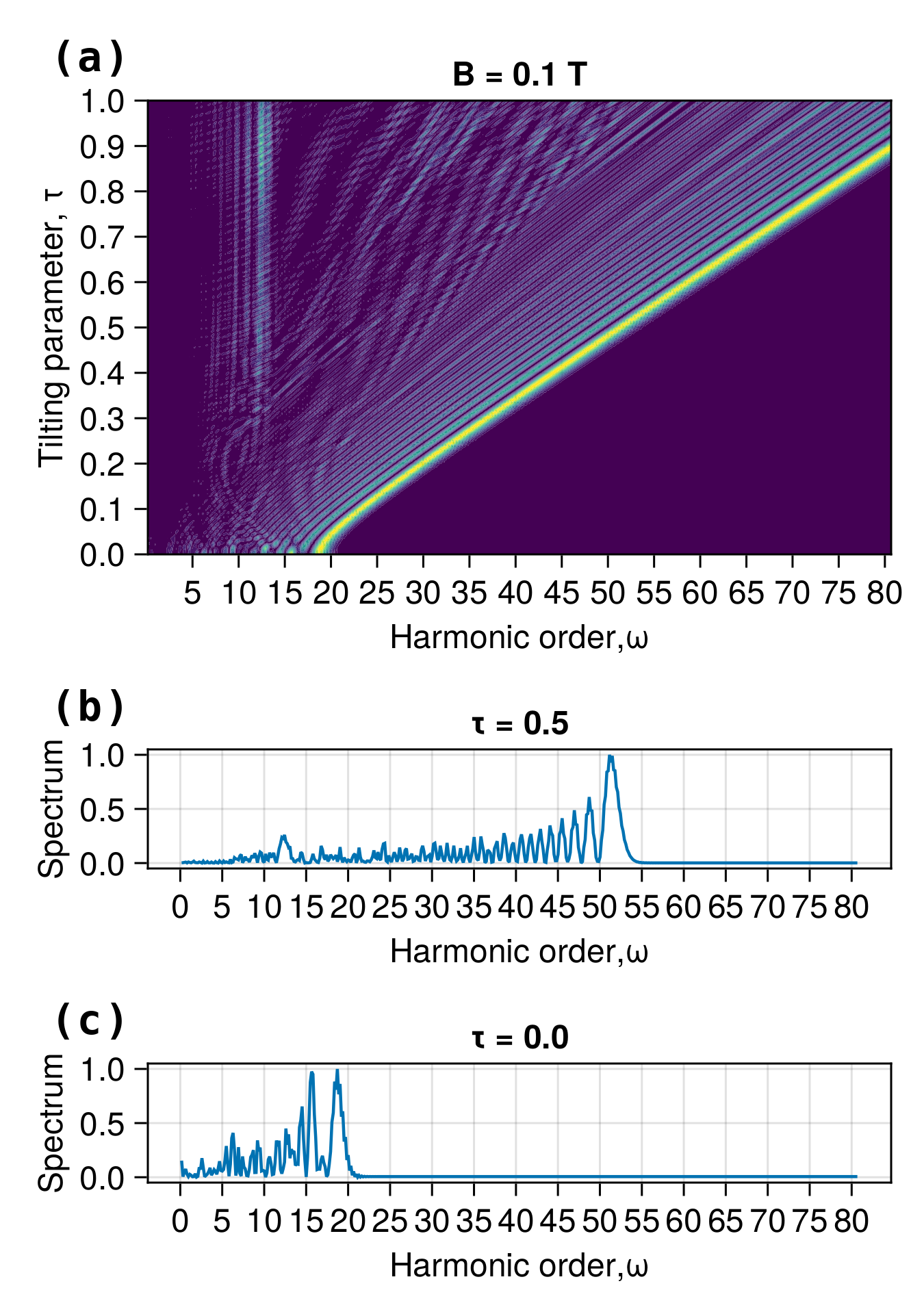}
    \caption{(a) Spectrum for different values of tilting parameter $\tau$, with the magnetic field intensity set to $B=0.1$T. Panels (b) and (c) report the individual spectrum for different values of $\tau-0$,and $\tau=0.5$, respectively. As it can be seen, the tilting of a cone shifts higher harmonic of the spectrum in higher orders, with a slope, estimated from panel (a), of about 72 Hz.}
    \label{fig:spectrum_tau_01}
\end{figure}
A schematic representation of the different set of selection rules for the untilted and tilted case is shown in Fig. (\ref{fig:landau_levels}). A more careful analysis of these selection rules reveals, that while the simplest model to describe light-matter interaction for untilted Dirac cones is a 3-level system (see, for example, Ref. \citenum{mhi}), in the case of tilted cones, the simplest system that catches the essential physics of tilted cones in a magnetic field is a 5-level system.
\begin{figure}[!t]
    \centering
    \includegraphics[width=0.8\linewidth]{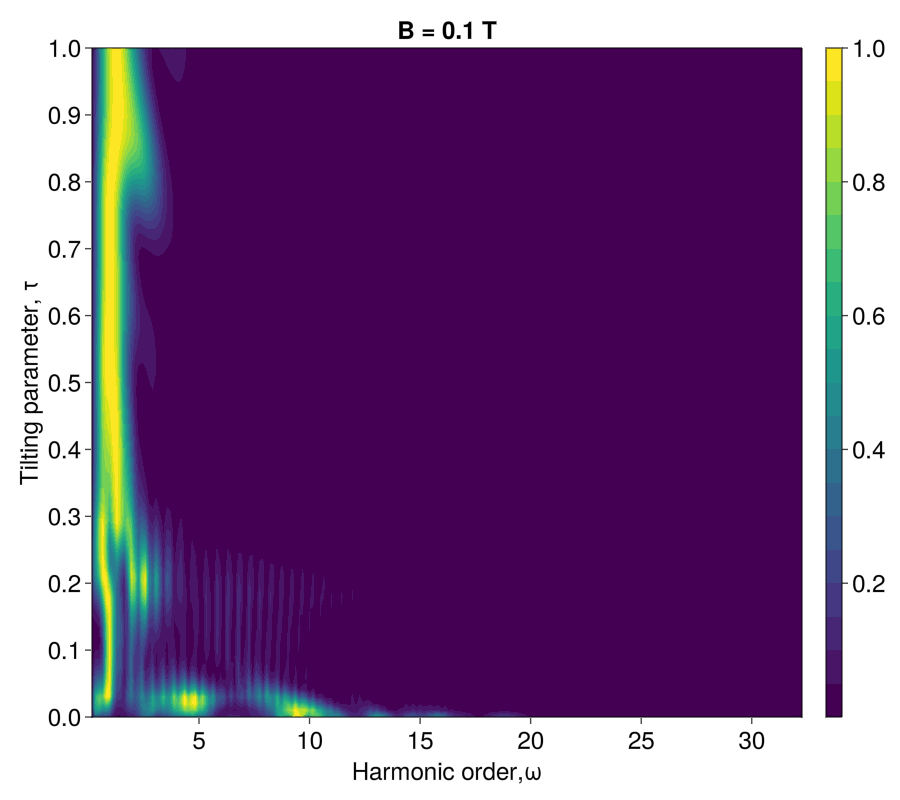}
    \caption{Spectrum for different values of tilting parameter $\tau$ with the magnetic field intensity set to $B=0.1$T. The polarisation of the impinging electromagnetic field in this case is orthogonal to the tilting direction, i.e., the electric field is polarised along the $\eta-$ ($y$-)direction. As it can be seen, the only effect of the tilting in this case is to suppress higher harmonics, than the fundamental, for values of the tilting parameter $\tau\geq 0.3$.}
    \label{fig:spectrum_tau_01_an_po}
\end{figure}
\begin{figure*}[!t]
    \centering
    \includegraphics[width=\textwidth]{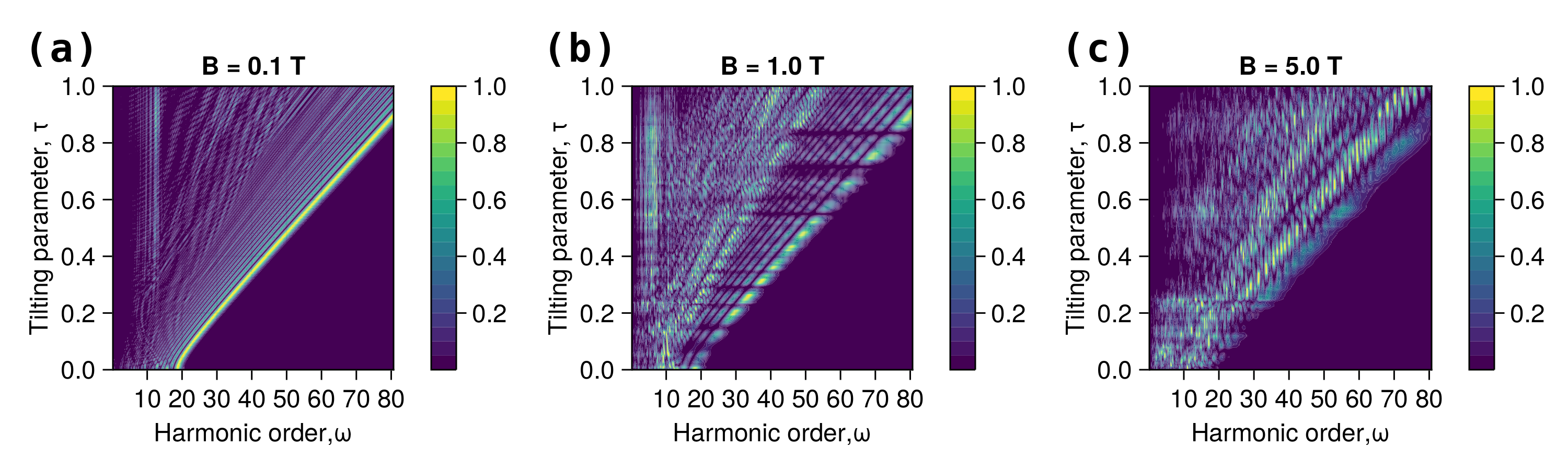}
    \captionof{figure}{Contour plots of spectrum with Harmonic order in $x$ axis, and tilting parameter in $y$ axis for different values of magnetic field amplitude: $0.1 T$, $1.0 T$ and $5.0 T$  figures (a),(b) and (c) respectively. We can see a trend on highest harmonic that a generated relative to tilting parameter and moreover this trend is conserved for different magnetic field amplitudes. Increase in magnetic field cause higher nonlinear effects thus more harmonics are generated in nonlinear optical response.}
    \label{fig:spectrum_combined}
\end{figure*}
\section{Nonlinear optical response}\label{sec:nonlinear_optical_response}
Once the coupled mode equations have been solved, we can proceed in calculating the current, with its usual definition, i.e.,
\beq
\vett{J}(t)=\expectation{\Psi(t)}{\boldsymbol\sigma}{\Psi(t)},
\eeq
where $\boldsymbol\sigma=\sigma_x\uvett{x}+\sigma_y\uvett{y}$. Substituting the expansion \eqref{eq37} into the above equation we get, for the components of the current
\barr
\label{eq:components_of_current}
J_{\mu}(t)&=&\sum_{n,m}\Bigg[(c_m^+)^*c_n^+e^{i(\omega_m-\omega_n)t}\expectation{\psi_m^+}{\sigma_{\mu}}{\psi_n^+}\nonumber\\
&+&(c_m^+)^*c_n^-e^{i(\omega_m+\omega_n)t}\expectation{\psi_m^+}{\sigma_{\mu}}{\psi_n^-}\nonumber\\
&+&(c_m^-)^*c_n^+e^{-i(\omega_m+\omega_n)t}\expectation{\psi_m^-}{\sigma_{\mu}}{\psi_n^+}\nonumber\\
&+&(c_m^-)^*c_n^-e^{-i(\omega_m-\omega_n)t}\expectation{\psi_m^-}{\sigma_{\mu}}{\psi_n^-}\Bigg],
\earr
and the expectation values of the Pauli matrices over the perturbed states can be calculated up to order $\mathcal{O}(\tau)$ using the results from the previous section. From here, we can then calculate the Fourier transform of the induced current, and the correspondent nonlinear signal as $I(\omega) \thicksim | \omega\, \mathbf{J}(\omega) |^2$, where $\mathbf{J}(\omega)$ is the Fourier transform of the nonlinear current.

To study nonlinear optical response of system under consideration we solve coupled mode equations for the time-dependent coefficients defined by Eq. (\ref{eq:coupled_mode_equations}) using Julia package \emph{DifferentialEquations.jl} \cite{rackauckas2017differentialequations}, and then we compute the nonlinear electrical current using Eq. (\ref{eq:components_of_current}) and definition of nonlinear signal given above. 
As said at the end of the previous section, we employ a 5-level model to describe the nonlinear response of 2D Weyl materials, i.e., our model only contains 2 levels in the valence band, namely $\ket{\psi_{-2}}$ and $\ket{\psi_{-1}}$, and two levels in the conduction band, i.e., $\ket{\psi_{1}}$ and $\ket{\psi_{2}}$. The zero energy state $\ket{\psi_0}$, common to both Landau oscillators, constitutes the fifth level. As initial condition, we assume that only the lowest LL is occupied, i.e., $c_{-2}(0)=1$. Moreover, we  assume an impinging electromagnetic pulse with an amplitude of $E_{0} = 10^{7}$ V /m, a carrier frequency of $\omega_{L} = 78 $ THz, resonant with the transition $\ket{\psi_{-2}}\rightarrow\ket{\psi_0}$, and a pulse duration of $\tau = 50$ fs. We have performed simulations with varying tilting parameter for both the cases of fixed and varying magnetic field strength. 
The results of these simulations are depicted in Figs. \ref{fig:spectrum_tau_01} and \ref{fig:spectrum_combined}, respectively. 
However, before focusing on those results, let us first concentrate on Fig. \ref{fig:spectrum_tau_01_an_po}. There, we plot the nonlinear response of a 2D Weyl material, in the presence of a weak magnetic field, for the case, where the impinging electric field is polarised in the direction orthogonal to that of the tilt, i.e., for this figure we have assumed $A_{\xi}=0$ and set $A_{\eta}$ to be equal to Eq. \eqref{eq23}.The parameters for this simulation, moreover, are the same as those listed above. As it can be seen from Fig. \ref{fig:spectrum_tau_01_an_po}, for the case of orthogonally polarised (with respect to the tilt direction) pulses, the only effect of the tilting is to suppress higher harmonics, for values of the tilting parameter $\tau\geq 0.3$, and no particularly interesting dynamics occurs.

The situation, where the impinging field is aligned with the direction of the tilt, on theother hand, is profoundly different. Let us first discuss the case of fixed magnetic field. To start with, let us use the value $B=0.1$ T for the magnetic field. We will then investigate the influence of different values of the magnetic field amplitudes later in this section. . As it can be seen from panel (c), for $\tau=0$, i.e., untilted cones, our results are consistent with those obtained in Ref. \cite{mhi}  (compare panel (c), for example, with Fig. 6 (a) therein), even though in our case we obtain a similar result with a magnetic field approximately one order of magnitude lower, than the one used in Ref. \cite{mhi}. This could be simply imputable to the presence of a nonzero gap between valence and conduction band in 2D Weyl materials, which is absent in graphene. As soon as we turn on the tilt, we can see, from panel (a) of Fig. \ref{fig:spectrum_tau_01}, that an increase of the tilt parameter $\tau$ corresponds to a blue shift of the harmonic spectrum, which, in turn, allows the generation of higher harmonics, than the untilted case. From Fig. \ref{fig:spectrum_tau_01} (a), moreover, it can be seen how the blue shift of the maximum of the harmonic spectrum is approximately linear with $\tau$. This phenomenon is a consequence of the fact, that the kinetic momentum of an electron in the vicinity of a tilted cone is given by $\vett{p}=\vett{k}+(e/c)\vett{A}^{(g)}+(e/c)\vett{A}$. In this scenario, the vector potential $\vett{A}^{(g)}$ linearly displaces the electron momentum [see Eq. \eqref{eq7}]. The action of this linear displacement, combined with the action of the tilting on the interaction Hamiltonian, ultimately makes sure, that the blue shift experienced by the nonlinear spectrum is linear in $\tau$. As a concrete example of the possibility this offers for light-matter interaction engineering, in Fig. \ref{fig:spectrum_tau_01} (b) we explicitly point to the case $\tau=0.5$, where the maximum of the spectrum now sits around the 50th harmonic. Considering the input carrier frequency of $\omega_L=78$ THz, its 50th harmonics corresponds to $\omega_{50}=3900$ THz, i.e., $\lambda\simeq 480$ nm. By suitably engineering 2D Weyl materials to possess a tilt of $\tau=0.5$ (along the $x$-direction, in our case, but the same line of reasoning would hold for an anisotropic tilt), it would be then possible to realize a frequency converter, capable of implementing efficient conversion of light between the THz and the visible (blue, in this case) domain. 

Figure \ref{fig:spectrum_combined}, instead, shows how different values of the magnetic field affect the tilt-induced blue shift. As it can be seen, while the shift induced by the tilting parameter $\tau$ remains, essentially, unaltered by an increasing amplitude of the applied magnetic field, an increase of the latter changes nonlinear optical response of the system. In particular, it redistributes the energy between the various harmonic, as it can be seen, for example, in Fig. \ref{fig:spectrum_combined} (b), where intermediate harmonics get higher intensity, than the case with small magnetic field [see panel (a), for example]. Moreover, increasing the magnetic field even further leads to a more complicated scenario, as depicted in Fig. \ref{fig:spectrum_combined} (c), where the nonlinear response as a function of $\tau$ becomes much more complex. However, that the blue shift effect reported by our simulations is genuinely an effect of the tilting parameter $\tau$, as it persists independently on the value of the applied magnetic field, as it can be seen from Fig. \ref{fig:spectrum_combined}, where, despite the ``noise" introduced by the higher magnetic field, the dependence of the nonlinear response on the tilting parameter $\tau$ remains the same.
\section{Conclusion}\label{sec:conclusion}
In this work, we have studied how tilted Dirac cones in 2D Weyl materials are affected by the presence of an external (or artificial) magnetic field. We have derived the analytical expression of the LL and eigenstates for the case of tilted cones, using first order perturbation theory, and shown, that when 2D Weyl materials in the presence of magnetic field interact with electromagnetic pulses, the tilting induces an extra set of selection rules, which extends the usual ones to $0<\Delta|n|\leq 2$. We have then computed the nonlinear optical response of 2D Weyl materials immersed in magnetic fields and have noticed, that by controlling the tilt of their cones, it is possible to achieve a versatile and precise control of their nonlinear spectrum, and that by engineering Weyl materials with greatly tilted cones, it is possible to achieve high-harmonic generation up to the 80th harmonic for $\tau\rightarrow 1$. However, accounting a more physically feasible situation, we have discussed the case of $\tau=0.5$, where the 50th harmonic appears in the spectrum, thus enabling efficient transfer of energy between THz and UV radiation. In comparison, Ref. \cite{mhi} discusses the efficient THz-to-visible conversion of radiation in graphene in the presence of a magnetic field of 2T. Here, instead, not only the tilting can serve as a tuning parameter to enhance the generated harmonic, allowing access to different spectral regions, but also enables the same efficiency with much lower values of the magnetic field, roughly one order of magnitude lower than that used in Ref. \cite{mhi}.

Finally, we have studied the impact of the magnetic field on this process and concluded, that while the overall trend is not changed, a change in the magnetic field intensity corresponds to a redistribution of energy through the harmonics in the spectrum, thus resulting, in a more complex picture, that could offer some degree of control on the frequency conversion process in such materials.

Our results show that by suitably engineering 2D Weyl materials to possess specific tilting properties, and by controlling the applied magnetic field (for example through strain and bending), it is possible to realise efficient THz-to-visible frequency converters, or to employ such devices (and, in particular, their sensitivity to the magnetic field) for sensing applications, where, for example, a local change in the lattice structure (due, for example to the adsorption of a certain molecule) could vary (via artificial gauge) the intensity of the magnetic field, and therefore the response of the material.
\section*{Acknowledgements}
Y.T. and M.O. acknowledge the financial support of the Academy of Finland Flagship Programme, Photonics research and innovation (PREIN), decision 320165. Y.T. also acknowledges support from the Finnish Cultural Foundation, decision 00221008.
L D. M. V. acknowledges support from the European Commission under the EU Horizon 2020 MSCA-RISE-2019 programme (project 873028 HYDROTRONICS) and of the Leverhulme Trust under the grant RPG-2019-363.
\section*{Appendix A: Second Order Correction to the Landau Spectrum}b
In this Appendix, we briefly present the general form of the first nonzero correction to the LL, as a function of the perturbative tilt parameter $\tau$. As stated in the main text, it is fairly easy to show, that $\varepsilon^{(1)}=0$. Therefore, the first nonzero correction to the LL energy spectrum would be quadratic in $\tau$. To achieve this, we need to consider also second order contributions to Eq. \eqref{eq14}, i.e.,
\barr\label{eqA1}
&&\left(\hat{H}_0 + \tau\hat{V}\right)\left(\ket{\psi_n^{(0)}}+\tau\ket{\psi_n^{(1)}}+\tau^2\ket{\psi_n^{(2)}}\right)\nonumber\\
&=&\left(\varepsilon_n^{(0)}+\tau\varepsilon_n^{(1)}+\tau^2\varepsilon_n^{(2)}\right)\nonumber\\
&\times&\left(\ket{\psi_n^{(0)}}+\tau\ket{\psi_n^{(1)}}+\tau^2\ket{\psi_n^{(2)}}\right),
\earr
from which, the second order correction to the energy eigenvalues $\varepsilon_n^{(2)}$ can be calculated, using the first-order results discussed in Sect. \ref{sect3b}, together with the requirement, that
\beq
2\braket{\psi_n^{(0)}}{\psi_n^{(2)}}+\braket{\psi_n^{(1)}}{\psi_n^{(1)}}=0,
\eeq
to obtain the following expression for the second-order correction to the energy (the interested reader can check out the whole calculation in any standard Quantum Mechanics book, as, for example, Ref. \cite{messiah})
\beq
\varepsilon_n^{(2)}=\sum_{m\neq n}\frac{\left|\expectation{\psi_m^{(0}}{\hat{V}}{\psi_n^{(0)}}\right|^2}{\varepsilon_n^{(0}-\varepsilon_m^{(0)}},
\eeq
where, as before, $\hat{V}=\hat{H}_{tilt}/\tau$, and $\hat{H}_{tilt}$ is given by Eq. \eqref{eq13}. To calculate the argument of the modulus square of the numerator, we can use Eq. \eqref{eq18} to get
\barr
\left|\expectation{\psi_m^{(0}}{\hat{V}}{\psi_n^{(0)}}\right|^2&=&\left|\alpha_{n,n+1}\delta_{m,n+1}-\alpha_{n,n-1}\delta_{m,n-1}\right|^2\nonumber\\
&=&\left|\alpha_{n,n+1}\right|^2\delta_{m,n+1}\nonumber\\
&+&\left|\alpha_{n,n-1}\right|^2\delta_{m,n-1}.
\earr
Notice, that in going from the first to the second line of the expression above, we have used the fact, that the mixed term coming from the modulus square is proportional to $\delta_{m,n+1}\delta_{m,n-1}=\delta_{n-1,n+1}=0$, and therefore can be neglected. Using the result above, we can then write the second-order correction as follows:
\beq
\varepsilon_n^{(2)}=\frac{\left|\alpha_{n,n+1}\right|^2}{\varepsilon_n^{(0)}-\varepsilon_{n+1}^{(0)}}+\frac{\left|\alpha_{n,n-1}\right|^2}{\varepsilon_n^{(0)}-\varepsilon_{n-1}^{(0)}}.
\eeq
We leave to the reader the rather cumbersome task to find the explicit expression of $\varepsilon_n^{(2)}$ as an explicit function of the index $n$ and the energy scale $\lambda$. Per se, this task in not particularly difficult; one, in fact, only needs to substitute in the expression above the explicit expressions for the coefficients $\alpha{n,n\pm 1}$ as given by Eqs. \eqref{eq19}, and the expression for $\varepsilon_mn^{(0)}$ as given by Eq. \eqref{eq10}. The resulting expression is quite complicated, and cannot be easily simplified to a nice and compact form.

Finally, we can then write the explicit form of the dependence of the Landau energies on the tilting parameter $\\tau$ as the following, second-order accurate, form
\beq
\varepsilon_n=s_n\sqrt{\lambda^2+|n|}+\tau^2\frac{B_n}{[\varepsilon_n^{(0)}-\varepsilon_{n+1}^{(0)}][\varepsilon_n^{(0)}-\varepsilon_{n-1}^{(0)}]},
\eeq
where
\beq
B_n=\left|\alpha_{n,n+1}\right|^2[\varepsilon_n^{(0)}-\varepsilon_{n-1}^{(0)}]+\left|\alpha_{n,n-1}\right|^2[\varepsilon_n^{(0)}-\varepsilon_{n+1}^{(0)}].
\eeq

\bibliography{DBbib}
\end{document}